\documentclass[prd,aps,preprintnumbers,A4paper,nofootinbib,superscriptaddress]{revtex4}
\usepackage{graphicx}
\usepackage{slashed}
\usepackage{verbatim} 
\newcommand{\be}{\begin{eqnarray}}
\newcommand{\ee}{\end{eqnarray}}
\usepackage[colorlinks]{hyperref}
\usepackage{nameref}
\usepackage{cancel,bm} 
\usepackage{amsmath,amssymb}
\usepackage{mathtools}
\usepackage{float}
\usepackage{color}
\usepackage{leftidx}
\usepackage{booktabs}
\usepackage{latexsym}
\usepackage{revsymb}
\usepackage{multirow}
\usepackage{hypcap}
\usepackage{array}
\usepackage[english]{babel}
\usepackage{subcaption}
\newcommand{\sT}{{\scriptscriptstyle T}}

\def\d{\delta}

\def\({\left(}
\def\[{\left[}
\def\){\right)}
\def\]{\right]}

\def\ln{\hbox{ln}}

\newcommand{\OP}{\mathcal{O}}

\newcommand{\xB }{x_{\scriptscriptstyle B}}
\renewcommand{\d}{\mathrm{d}}

\usepackage{multirow}
\usepackage{caption}
\usepackage[normalem]{ulem}
\usepackage{float}
\usepackage{multirow}
\usepackage{subcaption}

\newcommand{\rajesh}{\color{green}}

\usepackage{cancel}

\begin{document}

\title{TMD evolution effect on $\cos2\phi$ azimuthal asymmetry in a back-to-back production of $J/\psi$ and jet at the EIC}
\author{Raj Kishore}
\email{raj.kishore@ehu.eus}
\affiliation{Department of Physics, University of the Basque Country, Leioa-48940, Spain}
\affiliation{Centre for Frontiers in Nuclear Science, Stony Brook University, Stony Brook, NY 11794-3800, USA}
\author{Asmita Mukherjee}
\email{asmita@phy.iitb.ac.in}
\affiliation{Department of Physics, Indian Institute of Technology Bombay, Mumbai-400076, India}
\author{Amol Pawar}
\email{194120018@iitb.ac.in}
\affiliation{Department of Physics, Indian Institute of Technology Bombay, Mumbai-400076, India}
\author{Sangem Rajesh}
 \email{sangem.rajesh@vit.ac.in}
 \affiliation{Department of Physics, School of Advanced Sciences, Vellore Institute of Technology, Vellore,
Tamil Nadu 632014, India}
%
\author{Mariyah Siddiqah}
\email{shah.siddiqah@gmail.com}
\affiliation{Department of Physics, University of Kashmir, Kashmir, India}

\date{\today}
\begin{abstract}  

A back-to-back semi-inclusive $J/\psi + jet$ production is a promising process to study gluon transverse momentum distribution (TMDs) at the future electron-ion collider (EIC). A back-to-back configuration allows a higher transverse momentum for $J/\psi$. We present an extension of a previous work where we studied $\cos2\phi$ azimuthal asymmetry within the TMD factorization framework  for this process. We present  and compare the effect of TMD evolution on the asymmetry,  in two approaches that differ in the parameterization of the perturbative tails of the TMDs and the non-perturbative factors. We  show that the asymmetry depends on the parameterizations of the non-perturbative Sudakov factors in the larger $b_T$ region and on the perturbative part of the evolution kernel. We use NRQCD to estimate the $J/\psi$ production and show the effect of using different long-distance matrix element (LDME) sets. Overall, the asymmetry after incorporating TMD evolution is small, but increases with the transverse momentum imbalance of the $J/\psi$-jet pair. 

\end{abstract}
\maketitle
 
\section{Introduction}

 Over several decades, experiments on deep inelastic scattering(DIS) of electron or muon beams
off nucleons have provided valuable insights into how quarks and gluons (collectively called partons) share
the momentum of a fast-moving nucleon.  Various observables sensitive to these parton's transverse momentum have proven to be valuable tools to test the QCD dynamics at high-energy colliders.  Of particular interest is the semi-inclusive deep inelastic scattering process, where a hadron is observed in the final state, in addition to the scattered lepton. Depending on the transverse momentum of the hadron, one can calculate the cross-section using a collinear or TMD factorized framework. In the large transverse momentum region, collinear factorization is applicable, which involves collinear parton distributions and fragmentation functions.  If the transverse momentum is much smaller than the hard scale  $Q$, then one applies the TMD factorization \cite{Collins:2011zzd,Echevarria:2011epo,Echevarria:2012js}  and the observables can  be calculated in terms of the transverse momentum dependent (TMD) parton distribution functions (PDFs) and fragmentation functions(FFs),  collectively called TMDs in short. Generally, TMDs are non-perturbative and hence need to be extracted from the data.  In addition to the parton’s longitudinal momentum fraction $(x)$, they also give information about the nucleon's intrinsic transverse momentum;  thus providing an open window for exploration of the three-dimensional structure of hadrons in terms of their QCD elementary constituents. Recently, a unified description of TMD factorization has been proposed in \cite{PhysRevD.109.034035}  bridging the large and small $x$ regions. TMDs have attracted   enormous interest,  experiments at different facilities around the world such as the JLAB 12 GeV and the Relativistic Heavy Ion Collider (RHIC) BNL have dedicated programs to probe the TMDs. The upcoming electron-ion collider (EIC) will probe the quark and gluon TMDs over a large kinematical region. TMDs significantly expand the non-perturbative information \cite{Mulders_1996} on hadron structure including transverse momentum and polarization degrees of freedom as compared to conventional integrated  collinear PDFs. The majority of experimental data for the extraction of the TMDs come from the semi-inclusive deep inelastic scattering (SIDIS)\cite{Boer:1997nt,Bacchetta_2007} and Drell-Yan(DY)\cite{Tangerman:1994eh,Arnold:2008kf} processes. In these processes, a final hadron with some transverse momentum or a pair of leptons is detected, carrying the signature of the transverse momentum originating from the partons within the proton. The most interesting observables in these processes are single-spin asymmetries (SSA) and azimuthal asymmetries.
Probing of TMDs has been proposed through  high-energy processes such as semi-inclusive deep inelastic scattering (SIDIS), Drell-Yan processes \cite{Aybat2011}. In SIDIS, the production of back-to-back hadron plus jet/photon(prompt) 
\cite{Kishore:2019fzb,DCRKAMSRJpsiphoton,Kishore:2022ddb,Bor:2022:EvolTMDShF}, $J/\psi$ plus pion \cite{Banu:2024ywv}, $D$-meson plus jet \cite{Banu:2023vid}, and dijet \cite{delCastillo:2020omr,Kang:2020xgk} at EIC kinematics
These processes have enabled the extraction of TMDs through the study of azimuthal asymmetries, which are sensitive to the transverse momentum and spin correlations of partons and hadrons. In Reference \cite{Echevarria:2024idp}, TMD factorization has been proven for the first time for the electroproduction of $J/\psi$ in the small transverse momentum region. In this context, the NRQCD approach breaks down, necessitating the replacement of NRQCD LDMEs with TMD shape functions.  Recent years have witnessed significant achievements in understanding quark   TMDs both experimentally and theoretically, while the gluon   TMD sector is still a domain where much more research and investigation are needed. Unlike the collinear PDFs, TMDs are not universal as they are process dependent \cite{BOER2003201}  
and their process dependence is due to gauge link (Wilson line). Gluon TMDs require two gauge links for their gauge invariance depending on the process  considered. At the leading twist, there are eight gluon TMDs. Among those, the gluon Boer-Mulders function, $h_{1}^{\perp g}$, has gained a lot of attention in recent years.
 $h_{1}^{\perp g}$, a time reversal even TMD, represents the density of linearly polarized gluons in an unpolarized proton. Until now no experimental extraction has been made for this function.  In recent years, quite a few theoretical studies have been done, in particular in quarkonium production processes, to probe the linearly polarized gluon TMDs  ~\cite{Boer:2011, 
Efremov:2017iwh,Kishore:2019fzb,DCRKAMSRJpsiphoton,Kishore:2022ddb,Bor:2022:EvolTMDShF,Boer:2012bt, Dunnen:2014eta, Mukherjee:2015smo, Mukherjee:2016cjw, Mukherjee:2016qxa, Scarpa:2019fol, DAlesio:2019qpk, Kishore:2021vsm}. These  TMDs generate a $\cos2\phi$ azimuthal asymmetries as well as affect the transverse momentum distribution of the final state.  At a small longitudinal momentum fraction, $x$, quarkonium mainly originates from gluon fusion, hence being sensitive to the gluon TMDs.

In our previous work in Ref. \cite{Kishore:2022ddb}, we studied a back-to-back $J/\psi$ plus jet pair production in unpolarized electron-proton collision. We found it a promising channel to probe linearly polarized gluon TMDs.  We obtained a sizable estimate of the asymmetry, using recent parametrizations of the gluon TMDs, given in  \cite{Bacchetta2020} and \cite{DCRKAMSRJpsiphoton,Kishore:2022ddb}  in  the kinematical region to be accessed at  the EIC. The asymmetry was about $15-20$ \% without incorporating scale evolution of the TMDs. In large transverse momentum region,  one can describe the cross-section of the process using collinear factorization framework \cite{CATANI1999143} in terms of the collinear PDFs and their scale evolution is governed by the DGLAP equations \cite{Dokshitzer:1977sg,Gribov:1972ri,Altarelli:1977zs}. However, in the intermediate and low transverse momentum region, one uses TMD factorization.  The evolution of TMDs with the energy scale of the process, known as TMD evolution, is a critical aspect of understanding both perturbative and non-perturbative QCD effects. The Collins-Soper-Sterman (CSS) \cite{Collins:1981uw,Collins:1984kg,Collins:2011zzd} framework governs TMD evolution, describing how TMDs evolve with the hard scale of the process \cite{Collins:1988ig,Aybat2011,Scimemi2018,echevarria2015qcd,Echevarria:2014rua}. Perturbative components of TMD evolution, involving logarithmic resummations to account for large momentum transfers, are well-understood and can be calculated using perturbative QCD \cite{Aybat2011}. However, non-perturbative components, which dominate at low transverse momenta, require phenomenological modeling and introduce significant uncertainties\cite{Collins:2011zzd}. It is important to estimate the effect of TMD evolution on observables for a more realistic predictions \cite{Collins:2011zzd,Echevarria:2011epo,Echevarria:2012js,Bacchetta:2022awv}. In our previous work, 
\cite{Kishore:2022ddb} it was found that the asymmetry decreases after TMD evolution is incorporated. In this work, we investigate the effect of evolution  on the asymmetry in detail.

Several approaches have been adopted in the literature to parameterize TMD evolution, each with its own treatment of the perturbative and non-perturbative factors, a brief discussion can be found in \cite{Collins_2015}. These approaches differ in how they handle the Sudakov factors, which rescale TMDs to account for the effects of soft gluon emissions. Notable frameworks include the CSS formalism, the generalized TMD evolution framework proposed by Aybat and Rogers \cite{Aybat2011}, and the non-perturbative parameterizations by \cite{Aybat2011,Scarpa:2019fol,Bor:2022:EvolTMDShF}. Each of these frameworks offers a different perspective on the interplay between perturbative and non-perturbative effects.
We consider the CSS framework here, and we employ two different approaches to parameterize the perturbative tails and non-perturbative factors of TMDs. In one approach, evolution is considered at leading order in $\alpha_s$ as discussed in Refs. \cite{Boer:2020bbd,Boer_2023}. This approach is adopted to find the matching coefficient 
by comparing cross-section calculations in the TMD factorization and the collinear factorization frameworks at an intermediate transverse momentum. In another approach, we describe the perturbative tail of TMDs by considering only leading order terms \cite{Aybat2011,echevarria2015qcd,Echevarria2016}.  Non-perturbative factors are largely unknown, however, some parameterizations have been considered in Refs. \cite{Aybat2011,Scarpa:2019fol,Bor:2022:EvolTMDShF} and comparing them shows sensitivity to the estimates of asymmetry heavily. Understanding the sensitivity of azimuthal asymmetries to these different parameterizations is essential for refining theoretical models and improving the extraction of TMDs from experimental data.  Here, we systematically investigate the effect of these two TMD evolution approaches on the asymmetry.

 The rest of this paper is organized as follows. The analytical framework of our calculation is
discussed in Section \ref{sec2}. In Section \ref{sec3}, we summarize the two TMD evolution approaches. In Section \ref{sec4}, we present numerical results and discussions and, finally, in Section \ref{sec5}, we conclude.

\section{Azimuthal asymmetry in \texorpdfstring{$J/\psi$}~ and jet production}\label{sec2}
In this section, we will briefly summarize  the calculations of $\cos 2\phi$ azimuthal asymmetry in the production of back-to-back $J/\psi$-jet pair   in unpolarized electron-proton collision process,  for more details of the calculations and the kinematics,, Ref. \cite{Kishore:2022ddb}  is referred. We consider 
\begin{equation}
e^-(l) + p(P) \to e^- (l') + J/\psi (P_{\psi}) + Jet (P_{j}) + X. \nonumber
\end{equation} 

The four momenta of each particle are presented within their respective parentheses. The target proton and the electron beam are both unpolarized. The virtual photon-proton center-of-mass (cm) frame is considered where they move along the $+z$ and $-z$ directions, respectively.  The kinematics here is defined in terms of two light-like vectors with the help of a Sudakov decomposition, here chosen to be the momentum $P(=n_{-})$ of the incoming proton, and a second vector $n(=n_{+})$, obeying the relations $n\cdot P = 1$ and $n_{+}^2=n_{-}^ 2 = 0$.  At the next-to-leading (NLO) order in $\alpha_s$,  both quark and  gluon channels   contribute for the production of $J/\psi$.  However, the quark channel contribution is insignificant compared to gluon channel \cite{DAlesio:2019qpk}.  Hence we consider only the gluonic subprocess $\gamma^\ast(q)+ g(k)\to J/\psi (P_\psi) +g(P_j)$.
 The 4-momenta of the proton $P$, virtual photon $q$ and incoming lepton $l$  are  written as below,
\be\label{pq4m}
P^\mu&=&n^\mu_-+\frac{M_p^2}{2}n^\mu_+\approx n^\mu_-\,,\nonumber\\
q^\mu&=&-\xB n^\mu_-+\frac{Q^2}{2\xB}n^\mu_+\approx -\xB P^\mu+(P\cdot q)n^\mu_+\,,\nonumber\\
l^\mu &=&\frac{1-y}{y}\xB n^\mu_-+\frac{1}{y}\frac{Q^2}{2\xB }n^\mu_
++\frac{\sqrt{1-y}}{y}Q\hat{l}^\mu_\perp\,,
\ee
where $M_p$ indicates the mass of the proton.  The inelasticity variable, $y = \frac{P\cdot q}{P\cdot l}$, represents the fraction of the electron's energy transferred to the photon. The Bjorken variable $\xB$ is defined as $\xB=\frac{Q^2}{2P\cdot q}$, with $Q^2$ being the squared invariant mass of the virtual photon which satisfies the relation $Q^2\approx \xB y s$, where $s$ is the cm energy of the electron-proton system. The 4-momenta of the  initial gluon ($k$), $J/\psi (P_{\psi})$ meson, and final gluon ($P_j$) can be written as,

\be\label{fv:pc}
k^\mu &\approx& x P^\mu +k^\mu_{\perp g}, \,\nonumber \\
P_{\psi}^\mu&=&z(P\cdot q)n_+^\mu + \frac{M_{\psi}^2+\bm {P}_{{\psi}\perp}^2} {2z P\cdot q} P^\mu+ P_{{\psi}\perp}^\mu\,\nonumber \\
P_{j}^\mu&=&(1-z)(P\cdot q)n_+^\mu + \frac{\bm {P}_{j\perp}^2}{2(1-z) P\cdot q} P^\mu+ P_{j\perp}^\mu\,,
\ee
where, $x$ and $k_{\perp g}$ are the longitudinal momentum fraction and intrinsic transverse momentum of the initial gluon.  $z=\frac{P \cdot P_{\psi}}{P\cdot q}$ represents the energy fraction carried by the $J/\psi$  from the virtual photon in the proton rest frame.  $J/\psi$ mass is represented with $M_\psi$.

The total differential scattering cross-section for the  unpolarized process: $e+p\to J/\psi + Jet+ X$ can be written as \cite{Pisano:2013cya}

\begin{eqnarray}
 d\sigma &=& \frac{1}{2s}\frac{d^3l'}{(2\pi)^32E_{l'}}\frac{d^3P_{\psi}}{2E_{\psi}(2\pi)^3}\frac{d^3P_{j}}{2E_{j}(2\pi)^3} \int dx~d^2k_{\perp g}~ (2\pi)^4 \delta^4(q+k-P_{\psi}-P_{j})\nonumber\\
 && \frac{1}{Q^4}~ L^{\mu\mu'}(l,q)~~\Phi^{\nu\nu'}_{g}(x,k_{\perp g}^2)~ ~\mathcal{M}_{\mu\nu}^{\gamma^\ast + g \to J/\psi + g} ~~\mathcal{M}_{\mu'\nu'}^{\dagger~ \gamma^\ast + g \to J/\psi + g }\, .\label{totsig}
 \end{eqnarray}

 The function $\mathcal{M}_{\mu\nu}$ represents  the scattering amplitude of  $\gamma^\ast +g \to J/\psi +g$ partonic subprocess.  The Fock states  that will contribute  in the calculation are $n= {^3}S_1^{[1,8]},{^1}S_0^{[8]},{^3}P_J^{[8]},$ with $J=0,1,2$. The leptonic tensor $L^{\mu\mu'}$  describes the electron-photon scattering and can be written as,
\begin{eqnarray}
 L^{\mu\mu'}&=e^2(-g^{\mu\mu'}Q^2+2(l^{\mu}l^{'\mu'}+l^{\mu'}l^{'\mu})),
\end{eqnarray}
where $e$ represents the electronic charge.  We have used the TMD factorized framework to calculate the cross-section of this process  \cite{DAlesio:2019qpk,Bor:2022:EvolTMDShF,Banu:2024ywv} in the kinematics where the outgoing $J/\psi$ and the gluon-induced jet are almost back-to-back.  The gluon correlator, $\Phi^{\nu\nu'}_{g}(x,{k}_{\perp g}^2)$ describes the gluon content of the proton.  At the leading twist, for the case of an unpolarized proton, it can be parameterized in terms of two TMD gluon distribution functions as \cite{mulders2001transverse}
\begin{eqnarray}
 \Phi_U^{\nu\nu'}(x,k_{\perp g}^2)=-\frac{1}{2x}\bigg\{g_{\perp}^{\nu\nu'}f_1^g(x,k_{\perp g}^2)-\left(\frac{k_{\perp g}^{\nu}k_{\perp g}^{\nu'}}{M_p^2}+g_{\perp}^{\nu\nu'}\frac{k_{\perp g}^2}{2M_p^2}\right)h_1^{\perp g}(x,k_{\perp g}^2)\bigg\}.
 \label{eq.5}
\end{eqnarray}
Here, $g_{\perp}^{\nu\nu'}=g^{\nu\nu'}-P^{\nu}n^{\nu'}/P\cdot n-P^{\nu'}n^{\nu}/P\cdot n$. The quantities $f_1^g(x,k_{\perp g}^2) $ and $h_1^{\perp g}(x,k_{\perp g}^2)$ represent the unpolarized and linearly polarized gluon TMD{\rajesh{s}}, respectively. In the  kinematical region considered, the outgoing particles are almost back-to-back in the transverse plane with respect to the line of collision of
virtual photon-proton system;  the differential scattering cross-section can be written as  \cite{DAlesio:2019qpk}

\begin{align}\label{eq:Un}
\frac{ d\sigma}
{d z\,d y\,d \xB\,d^2\bm{q}_{\sT}\,d^2\bm{K}_{\perp}} & =\mathcal{N}\bigg[\bigl(\mathcal{A}_{0}+\mathcal{A}_{1} \cos\phi_{\perp}+\mathcal{A}_{2} \cos2\phi_{\perp}\bigr)f_{1}^{g}(x,\bm{q}_{\sT}^{2})+\bigl(\mathcal{B}_{0}  \cos2\phi_{\sT}\nonumber\\
&\qquad\qquad+\mathcal{B}_{1}  \cos(2\phi_{\sT}-\phi_{\perp})+\mathcal{B}_{2}  \cos2(\phi_{\sT}-\phi_{\perp})+\mathcal{B}_{3}  \cos(2\phi_{\sT}-3\phi_{\perp})\nonumber\\
 &\qquad+\mathcal{B}_{4}  \cos(2\phi_{\sT}-4\phi_{\perp})\bigr)\frac{ \bm q_\sT^2  }{M_p^2} \,h_{1}^{\perp\, g} (x,\bm{q}_{\sT}^2)\bigg]\,,
\end{align}
where $\mathcal{N}=\frac{1}{16(2\pi)^{4}ys^2z(1-z)Q^{4}}$ is the kinematical factor.   $\phi_\sT$ and $\phi_\perp$  are azimuthal angles.  The various angular modulations are denoted as $\mathcal{A}_i$ for $i=0$ and $\mathcal{B}_j$ for $j=0,1,2$. In the above equation, we define the sum and difference of the transverse momentum of the final particles as follows,
\be\label{eq:newfv}
\bm q_{\sT}={\bm P}_{\psi\perp}+{\bm P}_{j \perp}, \quad
\bm K_T= \frac{{\bm P}_{\psi\perp}-{\bm P}_{j \perp}}{2}\,.
\ee The cross-section in Eq.~\eqref{eq:Un}  has azimuthal modulations by which one can extract the specific gluon TMD by measuring the weighted azimuthal asymmetry.

\section{ TMD evolution }\label{sec3}
The study of evolution gives us insight into different aspects of the confined motion of partons. Just like the PDFs, TMDs also depend on the scale of the probe. For collinear PDFs, DGLAP evolution can be seen as the resummed series $\left[\alpha_s \ln (Q^2/\mu^2)\right]^n$, and the evolved kernel is purely perturbative. For TMDs, there is an additional scale $k_{\perp}$, which introduces additional terms $\left[\alpha_s \ln^2 (Q^2/k_{\perp}^2)\right]^n$, which can be non-perturbative when $k_{\perp}\sim \Lambda_{\text{QCD}}$. This gives us an important aspect of studying QCD in a non-perturbative regime.
From the  QCD field theory definitions of the TMD arising from the TMD factorization, it becomes evident that TMDs depend on two auxiliary scales: a renormalization scale $\mu$, which arises from renormalizing the
UV divergence, which separates high and low energy/mass scales from one another, and a rapidity scale, $\xi$, which regulates the rapidity divergences, separating soft and collinear momentum regions from one another \cite{aybat2012qcd, Scimemi2018,Collins:2011zzd}. The evolution in these scales is governed by the renormalization group and the Collins-Soper equations. 

 It is important to consider the scale evolution of TMDs to give predictions in the energy range of the present and future colliders.  It becomes easier to implement the TMD evolution in the impact parameter space, $\bm b_T$-space, where $\bm b_T$ is the conjugate of the transverse momentum $\bm q_T$.  
 The solution of the evolution equations from some initial scales 
  $\mu_b,~\xi_b$ to some final scales $\mu,~$ $\xi$ can be given by (see \cite{Scimemi2018} for a recent detailed analysis of two-scale evolution of TMDs):

\begin{equation}\label{eq:tmdevol}
\hat{F}(x, \bm{b}_T^2; \xi, \mu) = e^{-\frac{1}{2}S_A(b_T; \xi, \xi_b, \mu, \mu_b)} \hat{F}(x, \bm{b}_T^2; \xi_b, \mu_b),
\end{equation}
where $S_A$ is the perturbative Sudakov factor which resums the large logarithms of the type $ln(b_T\mu)$. It is spin-independent, thus the same for unpolarized and polarized TMDs.  The $\hat{F}(x,\bm b_T^2,\xi,\mu)$ represents  any kind of TMD. In the perturbative domain: $|b_T|<<1/\Lambda_{QCD}$, the TMDs contain perturbative information, which is absorbed in the Wilson coefficient. Using the operator product expansion (OPE) of the TMDs onto collinear functions, the TMDs at the initial scale can be written as
\cite{Collins:2011zzd,echevarria2014qcd,aybat2012qcd,Echevarria:2014rua}
\begin{equation}
\hat{F}(x,\bm{b}_T^2,\xi_b,\mu_b) =  \frac{1}{2\pi} \sum_{a=q,\bar{q},g} C_{g/a}(x,\xi_b,\mu_b)\otimes f^a_1(x,\mu_b) ,\label{TMDevol1}
\end{equation} 
where $f^a_1(x,\mu_b)$ is a collinear PDF. The above OPE holds only in the small $b_T$ region.  In this work, we only consider the gluon-initiated channel, and so the evolution of gluon TMDs. Here the perturbative Wilson coefficient function depends on the particular type of TMD considered, which can be expressed as a series expansion in powers of  $\alpha_s$ as follows:
\begin{align}
{C}_{g/a}(x, \mu_b) =  \delta_{ga}\delta(1-\hat{x})+\sum_{k=1}^\infty  \left (\frac{\alpha_s(\mu_b)}{2\pi} \right )^k C^{[k]}_{g/a}(x,\mu_b)\,,
\label{Coeff}
\end{align}
where the coefficient $C^{(k)}_{g/a}$ for unpolarized gluon TMD is given 
\begin{eqnarray}
C^{[1]}_{g/g}=-\frac{\pi^2}{12}\delta(1-\hat{x})\nonumber\\
C^{[1]}_{g/a}=-C_F\hat{x}\,,\label{coeff_f1}
\end{eqnarray} 
while for linearly polarized gluon TMD
\begin{eqnarray}
C^{[1]}_{g/g}=\frac{\alpha_s}{\pi}~C_A~\Bigg(\frac{\hat{x}}{x}-1\Bigg)\nonumber\\
C^{[1]}_{g/a}=\frac{\alpha_s}{\pi}~C_F~\Bigg(\frac{\hat{x}}{x}-1\Bigg)\, .\label{coeffh_1}
\end{eqnarray} 
The initial scale of the TMDs can be considered as $\mu_b\sim \sqrt{\xi_b}\sim b_T^{-1}$, to minimize the large logarithm in the coefficient functions\cite{Scimemi2018}. It is typically taken as, $\mu_b=b_0/b_T$, where $b_0=2e^{-\gamma_{E}},\gamma_E\approx 0.577$. 

 As discussed in the introduction, in this article we'll investigate the effect of TMD evolution and the approach for TMD evolution on the $cos ~2 \phi$ azimuthal asymmetry in back-to-back production of $J/\psi$ and jet \cite{Kishore:2022ddb}.  To study this, we need to evolve two TMDs, namely the unpolarized and linearly polarized gluon TMD. We use two approaches \cite{Bor:2022:EvolTMDShF,Boer:2020bbd} for TMD evolution, in the context of this article we call these Approach-A \cite{Boer:2020bbd} and Approach-B  \cite{Bor:2022:EvolTMDShF}, respectively. These two approaches differ in the way the perturbative Sudakov factor, $S_A$, has been parameterized, the expansion of the   coefficient function and the parameterization of the non-perturbative Sudakov factor.  A brief review of these two TMD evolution approaches is given below.


\subsection{Approach-A}
In  the approach-A, we expand the perturbative part of TMDs, Eq.~\eqref{eq:tmdevol}, to a fixed order in $\alpha_s$. This approach is motivated to find matching coefficients at different perturbative orders by comparing the cross section computed in the TMD factorization scheme to the cross section computed in the collinear factorization scheme at some intermediate scale. More details can be found in \cite{Boer:2020bbd}. 
The LO expression for $S_A$ in Eq.~\eqref{eq:tmdevol} is given by, 

\begin{eqnarray}\label{eq:Sud}
S_A(\bm b_\sT^2, \mu^2) =\frac{C_A}{\pi}\int_{\mu_b^2}^{\mu^2}\frac{\d\bar{\mu}^2}{\bar{\mu}^2}\, \alpha_s(\bar{\mu})\left ( \ln \frac{\mu^2}{\bar{\mu}^2} - \frac{11- 2 n_f/C_A}{6} \right )\nonumber\\
 = \frac{C_A}{\pi}\, \alpha_s \left (\frac{1}{2}  \ln^2 \frac{\mu^2}{\mu_b^2}  -  \frac{11- 2 n_f/C_A}{6}\, \ln \frac{\mu^2}{\mu_b^2}  \right ) .
\end{eqnarray}
 In the above expression, Eq.~\eqref{eq:Sud}, both the scales of TMDs are equal to $\mu$. Moreover, the typical choice of the final scale is the hard scale, $Q$, in the process, $i.e~ (\mu,\zeta)=(Q, Q^2)$ which is justified by minimizing the large logarithms of $Q/\mu$ \cite{Scimemi2018}. In the second line,  the running coupling of $\alpha_s$ in the integration over the scale is neglected. We substitute Eq.~\eqref{eq:Sud} and TMD at input scale, Eq.~\eqref{TMDevol1}, into Eq.~\eqref{eq:tmdevol}, we obtain the perturbative part of $f_1^g$ at $\mathcal{O}(\alpha_s)$ as,

\begin{align}
\widehat{f}_1^g(x, \bm{b}_\sT^2; \mu^2) &= \frac{1}{2\pi}\bigg(f_1^g(x, \mu_b^2) - \frac{\alpha_s}{2\pi}\Bigg[\Bigg(\frac{C_A}{2} \ln^2\frac{\mu^2}{\mu_b^2} - \frac{11C_A - 2 n_f}{6} \ln\frac{\mu^2}{\mu_b^2}\Bigg)f_1^g(x, \mu_b^2) \nonumber \\
&\quad +  \sum_{i=q,\bar{q},g}\int_{x_A}^1 \frac{d\hat{x}}{\hat{x}}\hat{C}^1_{g/i}f_1^i\left(\frac{x_A}{\hat{x}}, \mu_b^2\right)\Bigg],
\label{eq:f1g2}
\end{align}
where the coefficients, $\hat{C}^1_{g/g}$ and $\hat{C}^1_{g/q}$ at the input scale $\mu_b$ are  given in Eq.~\eqref{coeff_f1}.
We will now proceed to discuss the linearly polarised gluon distribution function $h_1^{\perp g}(x,\bm{b}_T^2)$, which corresponds to the correlation between the linear polarization of gluon and its transverse momentum within the unpolarized proton. Its description requires a helicity flip, therefore, a gluon exchange. So, the perturbative beam functions for $h_1^{\perp g}$ can be computed in the same way as the perturbative tail of $f_1^{ g}$, with the key difference that its expansion begins at $\mathcal{O}(\alpha_s)$.  This is because, for the  linearly polarized gluon TMD, the first term from  Eq.~\eqref{Coeff} is zero \cite{Boer:2020bbd}  and hence the LO contribution is linear in $\alpha_s$.  The coefficient functions at LO for $h_1^{\perp g}(x,\bm{b}_T^2)$ are given in Eq.~\eqref{coeffh_1} \cite{sun2011gluon,echevarria2015qcd}.

The  perturbative part of $\hat{h}_1^{\perp g}(x,\bm{b}_T^2)$ at $\mathcal{O}(\alpha_s)$ can be written as\cite{Boer:2020bbd},
\begin{eqnarray}
\hat{h}_1^{\perp g(2)}(x,\bm{b}_T^2;\mu^2)=&&\frac{\alpha_s}{2\pi^2M_p^2}\frac{1}{b_T^2}\Bigg[C_A\int_x^1\frac{d\hat{x}}{\hat{x}}\Bigg(\frac{\hat{x}}{x}-1\Bigg)f_1^g(\hat{x},\mu_b^2)\nonumber\\
&&~~~~~~~~~~~~~~~~~~~~+C_F\sum_{a=q,\bar{q}}\int_x^1\frac{d\hat{x}}{\hat{x}}\Bigg(\frac{\hat{x}}{x}-1\Bigg)f_1^a(\hat{x},\mu_b^2)\Bigg].\label{h1perpbt}
\end{eqnarray}
Now,  we could  write the  expressions of the  unpolarised and linearly polarised  gluon TMDs in the $q_T$ space by making one-to-one correspondence between the functions in impact parameter and  momentum space using  Fourier transformation \cite{tangerman1995intrinsic,van2016quark, Echevarria:2015uaa} 
\begin{eqnarray}
    \hat{F}^{(n)}(x,\bm{b}_T^2,\mu^2)\equiv \frac{2\pi n!}{M^{2n}}\int_0^{\infty} d q_T~q_T\left(\frac{q_T}{b_T}\right)^nJ_n(q_Tb_T)\hat{F}(x,\bm{q}_T^2,\mu^2),
    \label{gen}
\end{eqnarray}
here, $n$ is the rank of function in $\bm{q_T}$-space.  The $q_T^n$ dependence in the above equation originates from the parametrization of the gluon correlator given in Eq.~\eqref{eq.5}. As can be seen from this expression,  the unpolarized gluon TMD has no intrinsic transverse momentum dependent factor in front of it. In contrast, the linearly polarized gluon TMD has a rank 2 tensor structure dependent on the transverse momentum of the initial gluon, which in our kinematics is equivalent to $q_T$ via the momentum conservation. Hence, for the linearly polarized gluon TMD, there will be $n=2$ in Eq.~\eqref{gen}. We 
have also used the general formula for Bessel integral \cite{kovchegov2012quantum}
\begin{eqnarray}
    \int_0^{\infty} dk~k^{\lambda-1} ~J_{\nu}(kx)=2^{\lambda-1} x^{-\lambda}\frac{\Gamma(1/2(\nu+\lambda))}{\Gamma(1/2(2+\nu+\lambda))},
\end{eqnarray}
where $J_{\nu}(z)$ is the Bessel function of the first kind of order $\nu$.

It is important to note that the Fourier transformation requires the entire
$b_T$ region. However, the applicability of the above perturbative expressions is confined to a restricted $b_T$ range; $b_0/Q<b_T<b_{T_{max}}$. For the larger $b_T$ region, the contributions come from the non-perturbative regime. On the contrary, for smaller $b_T$ values, $\mu_b$ exceeds  the  hard scale, and the evolution should stop. One of the popular ways to keep using the above perturbative expressions for the whole range of $b_T$ is to follow the $b^{\ast}_T$ prescription, where $b_T$ is replaced by a function of $b_T$ that freezes in both these limits.   In the literature, there are several $b_T^{\ast}$ prescriptions,  the one we have adopted in this approach is given below,  \cite{Collins:2016hqq}

\begin{eqnarray}
b_T^{*}(b_c(b_T))=\frac{b_c(b_T)}{\sqrt{1+\(\frac{b_c(b_T)}{b_{T_{max}}}\)^2}}
\end{eqnarray}
where,
\begin{eqnarray}
b_c(b_T)=\sqrt{b_T^2+\(\frac{b_0}{Q}\)^2}.
\end{eqnarray}
This ensures, $\mu_b=\frac{\it{b}_0}{b_T^{*}(b_c)}$ always lies between $b_0/b_{T_{max}}$ and $Q$  when $b_T\xrightarrow{} \infty$ and $b_T\rightarrow 0$ respectively. The value assigned to $b_{T_{max}}$ in the above equation more or less marks the line where we shift from the perturbative to the non-perturbative realm. For our analysis, we have chosen $b_{T_{max}}=1.5~\mathrm{GeV}^{-1}$.  Moreover, just substituting $b_T \to b_T^*$ in the perturbative expressions will not allow performing the integration over the whole range of $b_T$, as doing so will add the (full) contribution of the perturbative part for the whole range of $b_T$, however, the perturbative part only valid for small $b_T$. So, we consider a ``non-perturbative" factor that suppresses or smooths out the perturbative contribution in the large $b_T$ region.    The  non-perturbative Sudakov factor, $S_{NP}$, is introduced in Eq.~\eqref{eq:tmdevol} as 
\begin{equation}\label{eq:tmdevol_full}
\hat{F}(x, \bm{b}_T^2; \mu) = \hat{F}(x, \bm{b}_T^{*2};\mu_b)e^{-\frac{1}{2}S_A(b_T^*; \mu)} e^{-S_{NP}}.
\end{equation}
There is no such standard functional form for this non-perturbative Sudakov factor. It can only be extracted from the data. However, $S_{NP}$ is typically considered proportional to $b_T^2$ for all $b_T$. By its definition,   as the non-perturbative factor freezes out the perturbative contributions in the large $b_T$ region, it is usually constrained by two conditions: one of which is that it should be equal to $1$ for $b_T = 0$, and for large $b_T$, it should decrease monotonically and vanish, typically within the confinement distance. A value of $b_{t;lim}$ is defined from this non-perturbative Sudakov factor such that $e^{-S_{NP}}$ becomes negligible($\sim10^{-3}$) and a characteristic radius($\mathrm{r}=\frac{1}{2}b_{t;lim}$) is established based on this $b_{t;lim}$  which defines the extent to which interactions take place from the proton's center. The results of evolution cannot be uniquely predicted using evolution equations until the non-perturbative part is reliably extracted from the data. This non-perturbative Sudakov factor plays an important role in the evolution. For the numerical outcomes, we employed two different functional forms for the non-perturbative Sudakov factor. The first one, discussed here, is detailed in \cite{Scarpa:2019fol}, while the second will be addressed in approach-B.  
\begin{align}
S_{NP}(b_c(b_T)) = \frac{A}{2}~ \ln\left(\frac{Q}{Q_{NP}}\right) b_c^2(b_T), ~~~~~~\quad Q_{NP} = 1 \, \text{GeV}.
\end{align}
The parameter $A$ controls the width of the non-perturbative Sudakov factor for a particular $Q$. For our numerical calculations, we have evaluated the value of A at $Q=4.47$ GeV given in table  \ref{tabnonpert}  for three different values of $b_{t;lim}$ with convergence criteria of $e^{-S_{NP}} \approx 0.001$,
\begin{table}[H]
\centering
\begin{tabular}{ c|c|c } 
 \hline\hline
$ A (\mathrm{GeV}^2)$ & $ b_{t;lim}~(\mathrm{GeV}^{-1})$ & $ r~ (\mathrm{fm})$  \\ 
 \hline\hline
 2.2697727& 2 & 0.2   \\ 
 \hline
 0.57415574 & 4 & 0.4  \\ 
 \hline
 0.14395514& 8 & 0.8  \\ 
 \hline\hline
\end{tabular}
\hspace{1cm}
\caption{Value of parameter $A$ used in $e^{-S_{NP}}$  evaluated at $Q=4.47$ GeV.}
\label{tabnonpert}
\end{table}
 Now, we are in a position to write the 
unpolarised gluon TMD in the  $\bm{q_T}$-space as,
\begin{eqnarray}
f_1^g(x,\bm{q}_T^2;\mu^2) &=& \int^\infty_0 b_T ~db_T J_0(b_Tq_T) \nonumber\\
&&  \frac{1}{2\pi}\left(f_1^g (x, \mu_b^2)+\frac{\alpha_s}{2\pi}\left (({C}^{[1]}_{g/g} \otimes f_1^g) (x, \mu_b^2)+\sum_{a = q, \bar q} ({C}^{[1]}_{g/a} \otimes f_1^a) (x, \mu_b^2)-
\right.\right.\nonumber\\
&&\left.\left.\bigg[
 \frac{C_A}{2}  \ln^2 \frac{\mu^2}{\mu_b^2}  -  \frac{11C_A- 2 n_f}{6}\, \ln \frac{\mu^2}{\mu_b^2} \bigg]f_1^g (x, \mu_b^2) \right ) \right)e^{-S_{NP}(b_T,Q)}\,.
\end{eqnarray}

While the linearly polarized gluon TMD $h_1^{\perp g}$ in the $\bm{q}_T$-space can be written as,
\begin{eqnarray}
    \frac{q_T^2}{2M_p^2}h_1^{\perp g}(x,\bm{q}_T^2;\mu^2)=&& \frac{\alpha_s}{2\pi^2} 
    \int_0^{\infty} d {b_T}{b_T}~J_2(q_Tb_T)~\Bigg[C_A\int_x^1\frac{d\hat{x}}{\hat{x}}\Bigg(\frac{\hat{x}}{x}-1\Bigg)f_1^g(\hat{x},\mu_b^2)\nonumber\\ &&~~~~~+C_F\sum_{p=q,\bar{q}}\int_x^1\frac{d\hat{x}}{\hat{x}}\Bigg(\frac{\hat{x}}{x}-1\Bigg)f_1^p(\hat{x},\mu_b^2)\Bigg]e^{-S_{NP}(b_T,Q)}.\label{eq:finalh1perpg}
\end{eqnarray}


 \subsection{Approach-B}

 In approach-A discussed above, we expanded the resummed Sudakov exponent to a fixed order in $\alpha_s$, namely, we considered the perturbative part of TMDs up to $\mathcal{O}(\alpha_s)$.  In approach-B, we consider only the leading terms in the perturbative tails of both TMDs as defined by OPE in Eq.~\eqref{TMDevol1} at the initial scale, $\mu_b$ and consider the resummed Sudakov exponent to evolve the TMDs at the final scale, $Q$ \cite{Bor:2022:EvolTMDShF}. 
At LO, the perturbative tails of $\hat{f}_1^g$ is given by the collinear PDF as
\begin{eqnarray}
    \hat{f}_{1}^{g}(x,b_{T},\mu_b)&=f_{g/P}(x;\mu_b)+ \OP(\alpha_{s}) + \OP(b_{T}\Lambda_{\text{QCD}}) \hspace{1mm},
\end{eqnarray}
however, the perturbative tail of $\hat{h}_1^{\perp g}$ at LO starts at $\OP (\alpha_s)$ and is  given as, 
\begin{align}
\hat{h}_{1}^{\perp g}(x,b_{T},\mu_b) &=-\frac{\alpha_{s}(\mu_b)}{\pi}\int_{x}^{1} \frac{dx'}{x'} \hspace{1mm} \bigg(\frac{x'}{x}-1\bigg)\hspace{1mm} \bigg\{ C_{A} \hspace{1mm}f_{g/P}(x';\mu_b)+ C_{F} \sum_{i=q,\bar{q}}  f_{i/P}(x';\mu_b) \bigg\} \nonumber \\ & + \OP(\alpha_{s}^{2}) + \OP(b_{T}\Lambda_{\text{QCD}}) \hspace{1mm}.\label{h1perptail}
\end{align}

Note that there is no collinear version of $h_{1}^{\perp g}$, also, one can see that both expressions are determined by the collinear distributions $f_{i/P}$, that  start at different orders in $\alpha_{s}$. This approach  has been  considered for TMD evolution both in pp \cite{BOER2014421,Scarpa:2019fol,Echevarria_2019,echevarria2014qcd,echevarria2015qcd,Aybat2011} and ep \cite{Mukherjee:2016qxa,Bor:2022:EvolTMDShF,Aybat2011} collisions.   For semi-inclusive production of $J/\psi$  in the TMD factorized framework, one needs to include TMD shape functions \cite{Echevarria_2019, Fleming_2020,Boer_2023,Bor:2022:EvolTMDShF}. The TMD shape functions incorporate the smearing effect due to the transverse momentum of soft gluon emission during the formation of the bound state. The shape function,  $\Delta^{[n]}$ is needed for a full TMD factorization of the quarkonium production process, particularly in the case of small transverse momentum, $q_T<<M_{\psi}$ and plays a crucial role in accounting for two soft processes: the formation of the quarkonium-bound state and the resummation of soft gluons. However, the shape functions can be process-dependent \cite{Boer_2023}, and so far, their functional form for the $J/\psi$-jet pair production process has not been calculated. In this work, we have not incorporated TMD shape function, and only used the standard sets of LDMEs for $J/\psi$ formation. 

 By including the one-loop running of  $\alpha_s$ at the leading order  and dropping the shape function, the perturbative Sudakov factor can be written as \cite{Bor:2022:EvolTMDShF}, 
\begin{align}
S_{A}(b_{T}; Q, \mu_{b}) = & -\frac{36}{33-2 n_{f}}\bigg[\text{ln} \hspace{1mm} \frac{Q^{2}}{\mu_{b}^{2}} +  \text{ln}\hspace{1mm} \frac{Q^{2}}{\Lambda_{\text{QCD}}^{2}}  \hspace{1mm} \text{ln} \hspace{1mm} \bigg(1- \frac{\text{ln} \hspace{1mm} (Q^{2}/\mu_{b}^{2})}{\text{ln} \hspace{1mm} (Q^{2}/\Lambda_{\text{QCD}}^{2})} \bigg) \nonumber \\
& + \bigg(\frac{11-2n_{f}/C_{A}}{6} \bigg)\hspace{1mm} \text{ln} \hspace{1mm} \bigg(\frac{\text{ln} \hspace{1mm} (Q^{2}/\Lambda_{\text{QCD}}^{2})}{\text{ln} \hspace{1mm} (\mu_{b}^{2}/\Lambda_{\text{QCD}}^{2})} \bigg)\bigg] + \OP(\alpha_{s}^{2})\hspace{1mm}.
\end{align}
It should be noted here that $S_A$ is spin-independent and thus the same for all the  TMDs. In \cite{Bor:2022:EvolTMDShF} a constant term coming from the shape function due to soft gluon emission is included in $S_A$, here we have omitted this term for our process.  The final scale of the evolution is the hard scale, $Q$ as considered in approach-A. However,  in this approach we consider a slightly different initial scale, $\mu_b'$ that ensures $b_0/Q \leq b_T$.  We also have a slightly different form of $b_T^*$ prescription \cite{BOER2014421}
\begin{align}
\mu_{b}\to \mu_{b}'=\frac{Qb_{0}}{Qb_T+b_{0}} \to \mu_{b^*}' = \frac{Qb_{0}}{Qb_T^*+b_{0}}\hspace{1mm},
\label{primestar}
\end{align}
\begin{align}
 b_{T}^{*}(b_{T}) = \frac{b_{T}}{\sqrt{1+\left(b_{T}/b_{T_{max}}\right)^{2}}} \hspace{1mm},
\label{bstar}
\end{align}
that ensures $b_T^{*} \leq b_{T_{max}}$ and $\mu_b'\leq Q$. In approach-B, we consider a different functional form of non-perturbative Sudakov factor, which is inspired by the parameterization obtained from fitting SIDIS, DY and Z-boson production data \cite{Aybat2011}, given as 
\begin{align}
S_{NP}(b_{T};Q) = \bigg[ A \hspace{1mm} \text{ln} \frac{Q}{Q_{NP}}+ B(x)\bigg]b_{T}^{2}\hspace{1mm},
\label{SNPABform}
\end{align}
here $Q_{NP}=1.6$ GeV. The values corresponding to the parameter $A$ can be found in Table \ref{tabnonpert1}. The $B$-term is associated with the intrinsic transverse momentum of the TMDs, which generally varies with $x$. To determine these values, we have applied Eq.~(41) from \cite{Bor:2022:EvolTMDShF} at $Q=Q_{NP}$. As previously discussed, the non-perturbative Sudakov factor should be unity when $b_T$ equals zero, and for large $b_T$ it should vanish, to eliminate contributions originating from regions significantly outside the proton. 
\begin{table}[H]
\centering
\begin{tabular}{ c|c|c } 
 \hline\hline
$ A (\mathrm{GeV}^2)$ & $ b_{t;lim}~(\mathrm{GeV}^{-1})$ & $ r~ (\mathrm{fm})$  \\ 
 \hline\hline
 0.80& 2 & 0.2   \\ 
 \hline
 0.20 & 4 & 0.4  \\ 
 \hline
 0.05& 8 & 0.8  \\ 
 \hline\hline
\end{tabular}
\caption{Values of the parameter A used in $e^{-S_{NP}}$ evaluated at $Q=12$ GeV. }
\label{tabnonpert1}
\end{table}
Considering all the above equations, one can finally write the TMDs in the $q_T$ space
\begin{align}
f_{1}^{g}(x,\bm{q}_{T}^{2},\mu)
&= \int_{0}^{\infty} \frac{db_{T}}{2\pi} \hspace{1mm} b_{T}\hspace{1mm} J_{0}(b_{T}q_{T})\hspace{1mm} \hat{f}_{1}^{g}(x,b_{T}^{*},\mu^\prime_{b^\ast})\hspace{1mm}e^{-S_{A}(b_{T}^{*};Q,\mu_{b^*}')} e^{-S_{NP}(b_{T};Q)}  \hspace{1mm},\nonumber\\
h_{1}^{\perp g}(x,\bm{q}_{T}^{2},\mu)&= - \hspace{1mm} \int_{0}^{\infty} \frac{db_{T}}{2\pi} \hspace{1mm} b_{T} \hspace{1mm} J_{2}(b_{T}q_{T})\hspace{1mm} \hat{h}_{1}^{\perp g}(x,b_{T}^{*},\mu^\prime_{b^\ast}) \hspace{1mm}  \label{eq:New_approach_TMD}\hspace{1mm} e^{-S_{A}(b_{T}^{*};Q,\mu_{b^*}')} e^{-S_{NP}(b_{T};Q)}.
\end{align}

\subsection{Comparing the two non-perturbative Sudakov factors}
\begin{figure}[H]
    \centering
    \begin{subfigure}{0.49\textwidth}
        \includegraphics[height=4cm,width=6.5cm]{ 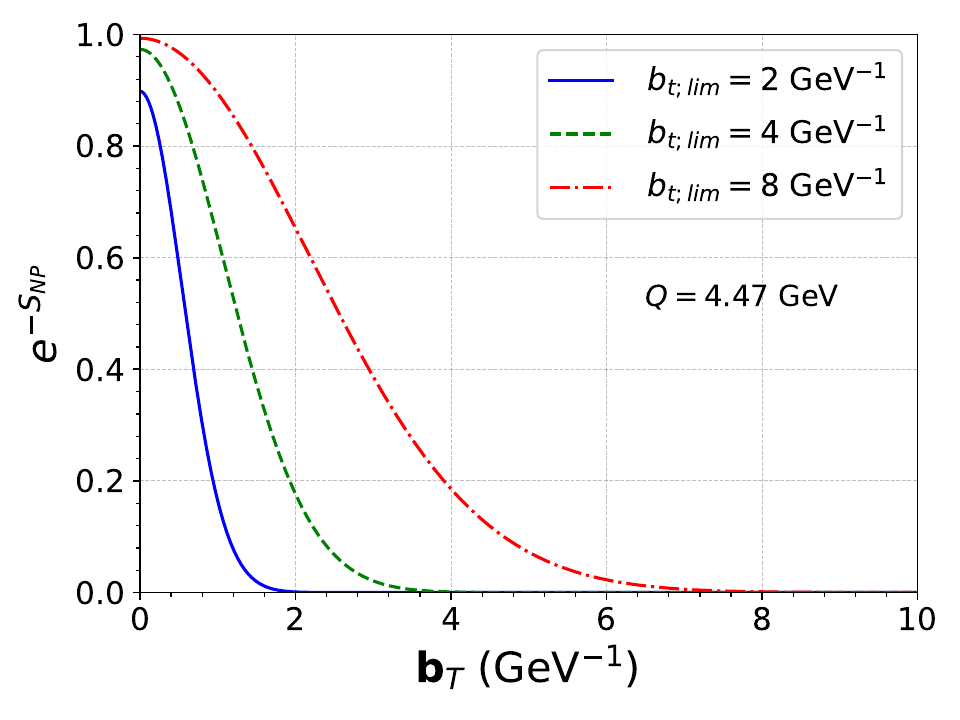}
        \caption{}
    \end{subfigure}
    \begin{subfigure}{0.49\textwidth}
        \includegraphics[height=4cm,width=6.5cm]{ 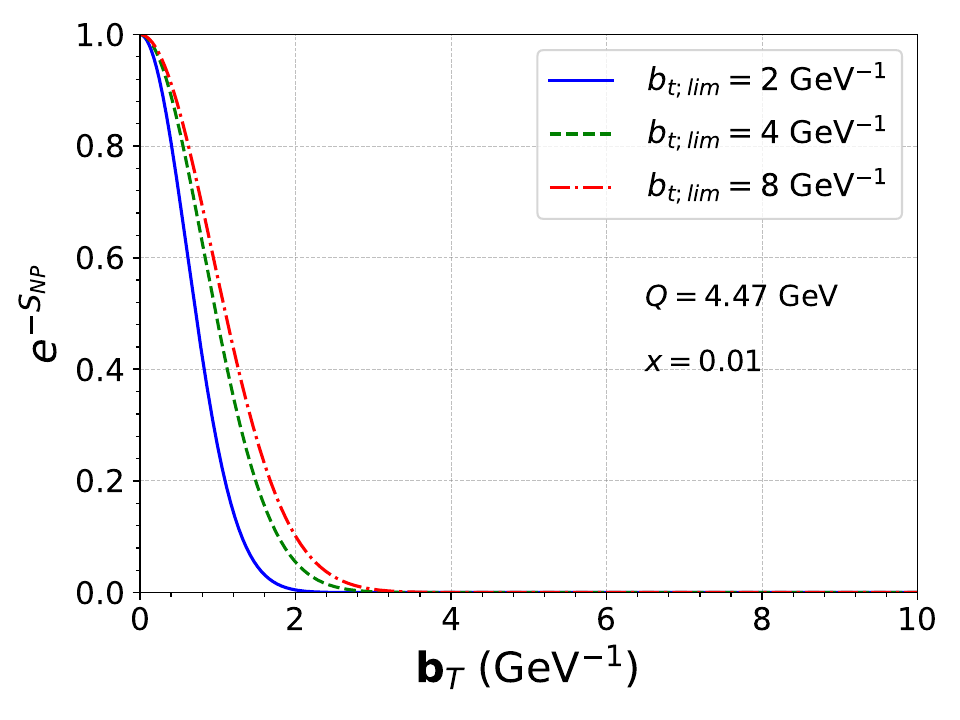}
        \caption{}
    \end{subfigure}
    \caption{Non-perturbative Sudakov factor $e^{-S_{NP}}$ plotted as a function of $b_T$ for $x=0.01$ at $Q=4.47~\text{GeV}$ (a) from Ref.~\cite{Scarpa:2019fol} (b) from Ref.~\cite{Bor:2022:EvolTMDShF}.}
    \label{fig:snp}
\end{figure}
As already said, the non-perturbative Sudakov factor $S_{NP} $  is not unique, since its exact functional form cannot be calculated. In the literature, different parameterizations have been used
depending on the type of TMDs. The basic idea starts with the fact that it should suppress the
contribution of the perturbative part of the TMDs when going to the higher $b_T$ domain (a non-perturbative region). A general constraint on $S_{NP}$ can be defined as follows: the value of $e^{-S_{NP}}$ should approach 1 as $b_T \rightarrow 0$ and vanish for large $b_T$, typically within the confinement distance.
This also ensures the convergence of the results.  In the two approaches of the TMD evolution as discussed, we have used different parameterizations of $ S_{N P}$. Both parameterizations satisfy the general constraint, however, they are different in their respective slope of the curve. We have illustrated the difference in Fig.~\ref{fig:snp}  for two choices of $ S_{N P}$, from Ref.~\cite{Scarpa:2019fol} and Ref.~\cite{Bor:2022:EvolTMDShF} respectively. As seen in Fig. Fig.~\ref{fig:snp}(b), the functional form of $e^{(-S_{NP})}$ leads to 
a relatively larger suppression of the contributions from the perturbative part in the $b_T$ range from $b_{T_{max}}$ to roughly the size of a proton. However, in Fig.~\ref{fig:snp}(a), we have relatively more contributions in that $b_T$ range. This leads to a large band of uncertainties going from $b_{t;lim} = 2-8$ GeV$^{-1}$ in the approach-A as seen in Fig.~\ref{fig1}.
\section{Numerical results for  \texorpdfstring{$\cos2\phi_T$} ~~azimuthal asymmetry using two TMD evolution approaches}\label{sec4}

In this section, we present numerical estimates of the $\cos2\phi_T$ azimuthal asymmetry for $\gamma^*+g \rightarrow J/\psi + \mathrm{jet}$ in the TMD evolution framework. We have incorporated two different TMD evolution approaches, referred to as Approach-A and Approach-B, within the context of this article and compared the asymmetry resulting from 
both approaches.  As seen from the discussion in the previous section, the approaches differ slightly in the perturbative coefficient and the Sudakov factor, and the parametrization of the non-perturbative Sudakov factors. We consider small-$b_T$ operator product expansion for defining the perturbative tails of the TMDs, at the initial scale, in terms of integrated distributions. In approach-A, we expand the evolution kernel, $e^{-1/2S_A}$  at leading log in resummation,  to a fixed order in $\alpha_s$ and considered the perturbative part of TMDs up to $\mathcal O {(\alpha_s)}$. Whereas, in Approach-B, we consider only the leading order (LO) terms in the perturbative tails of TMDs multiplied with the evolution kernel, where in the exponent, we keep the leading order terms and include the effect of the running of $\alpha_s$.

\begin{figure}[H]
	\begin{center} 
	\begin{subfigure}{0.49\textwidth}
\includegraphics[height=4cm,width=6.5cm]{ 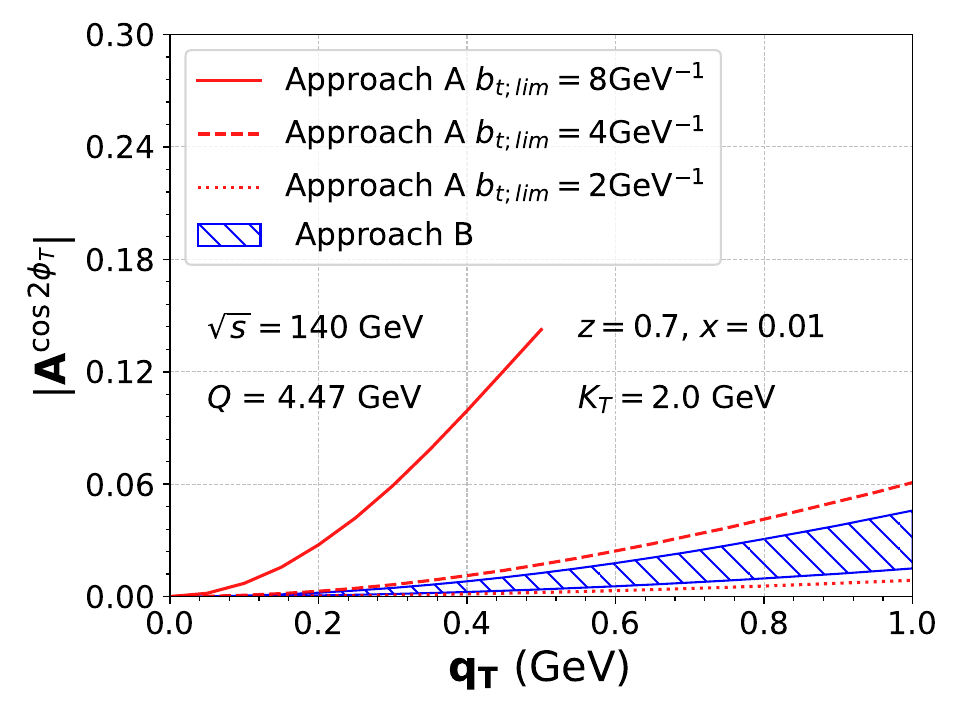}
  \caption{}
	\end{subfigure}
     \begin{subfigure}{0.49\textwidth}
\includegraphics[height=4cm,width=6.5cm]{ 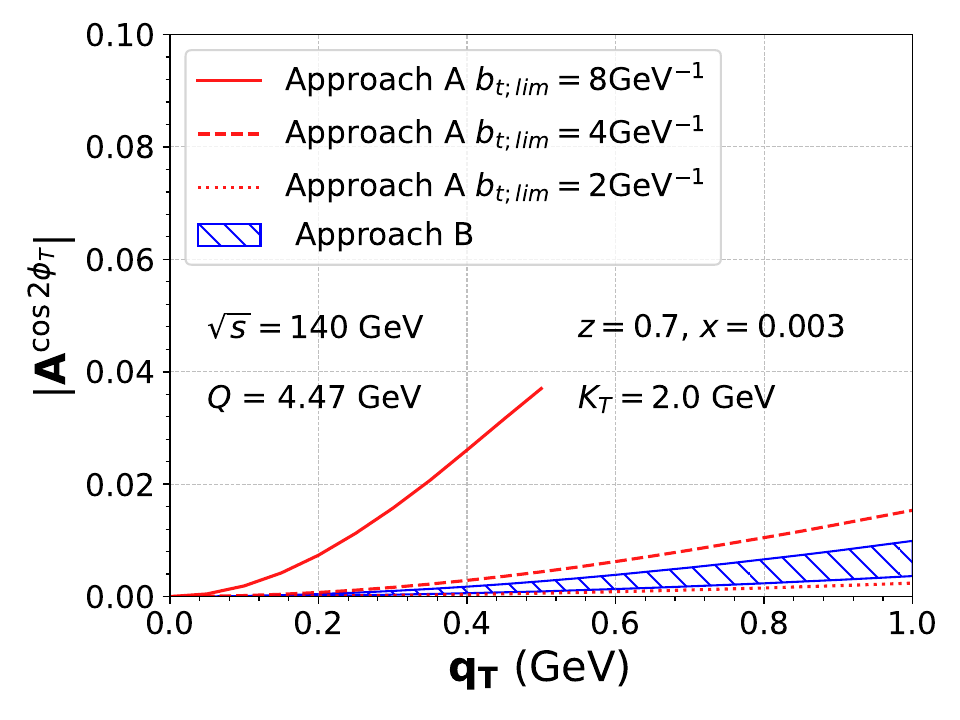}
	        \caption{}
	    \end{subfigure}
	\end{center}
\caption{\label{fig1} $\cos2\phi_T$ azimuthal asymmetry for Approach-A and -B in $e+p\rightarrow e +J/\psi+\mathrm{jet}+X$ as function of $q_T$ at fixed values of $K_T=2.0 ~\mathrm{GeV}$, $\sqrt{s}=140$ GeV, $z=0.7$, $Q=4.47$ GeV. For panel (a) $x=0.01$, while for panel (b) $x=0.003$ is chosen respectively. {The band is obtained by varying $b_{t;lim}\{2-8\}~\mathrm{GeV}^{-1}$ in approach-B.}}
\end{figure}

\begin{figure}[H]
	\begin{center} 
	\begin{subfigure}{0.49\textwidth}
\includegraphics[height=4cm,width=6.5cm]{ 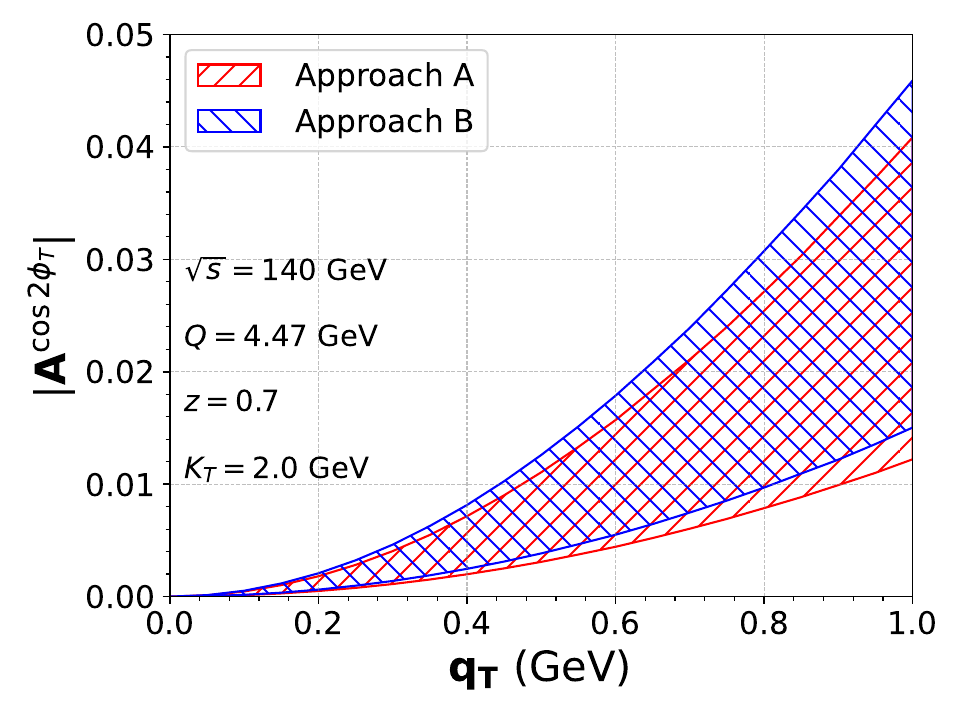}
  \caption{}
	\end{subfigure}
	    \begin{subfigure}{0.49\textwidth}
\includegraphics[height=4cm,width=6.5cm]{ 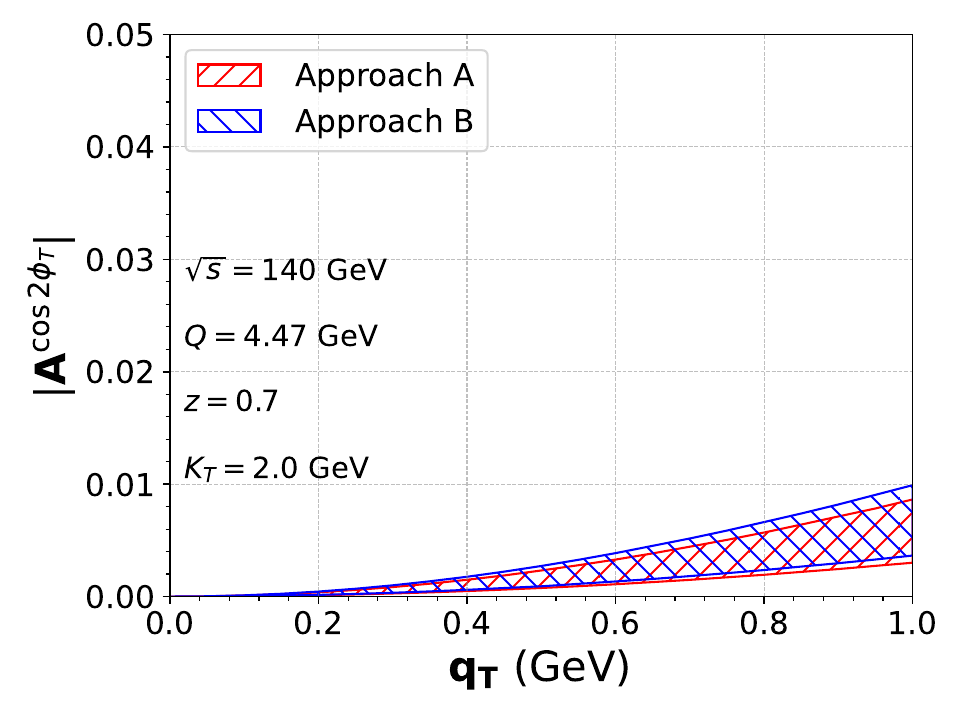}
	        \caption{}
	    \end{subfigure}
	\end{center}
\caption{\label{fig4} $\cos2\phi_T$ azimuthal asymmetry for Approach-A and -B in $e+p\rightarrow e +J/\psi+\mathrm{jet}+X$ as function of $q_T$ at fixed values of $K_T=2.0 ~\mathrm{GeV}$, $\sqrt{s}=140$ GeV, $z=0.7$, $Q=4.47$ GeV. For panel (a) $x=0.01$, and for panel (b) $x=0.003$ respectively. The bands shown in these figures are for $b_{t;lim}\{2-8\}~\mathrm{GeV}^{-1}$. The non-perturbative Sudakov factor is the same for both approaches given in Eq.~\eqref{SNPABform}.}
\end{figure}

\begin{figure}[H]
	\begin{center} 
	\begin{subfigure}{0.49\textwidth}
\includegraphics[height=4cm,width=6.5cm]{ 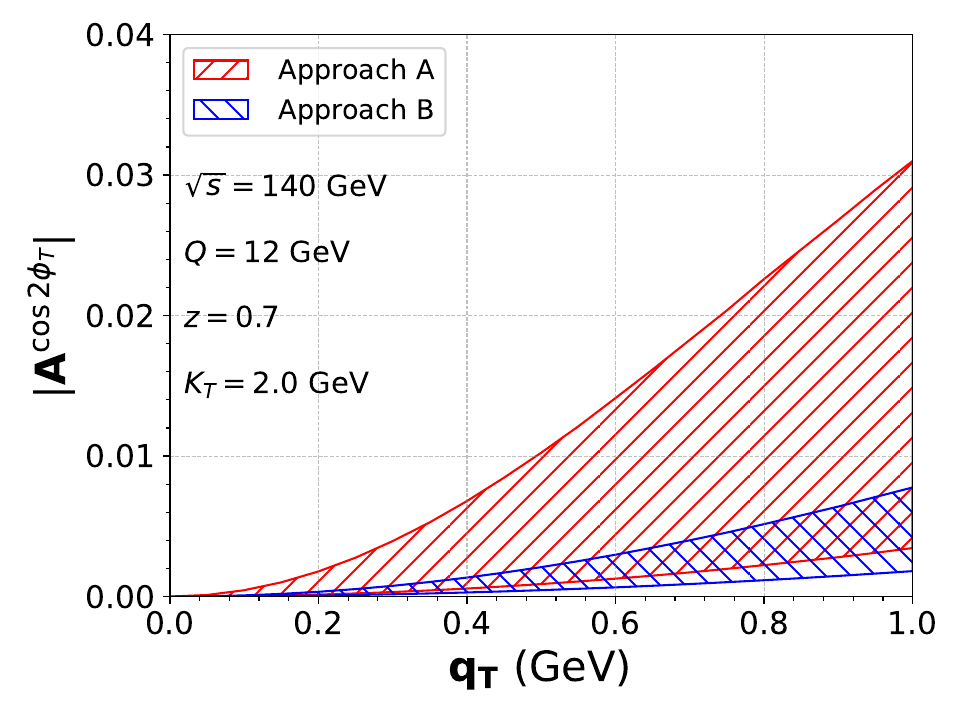}
  \caption{}
	\end{subfigure}
	    \begin{subfigure}{0.49\textwidth}
\includegraphics[height=4cm,width=6.5cm]{ 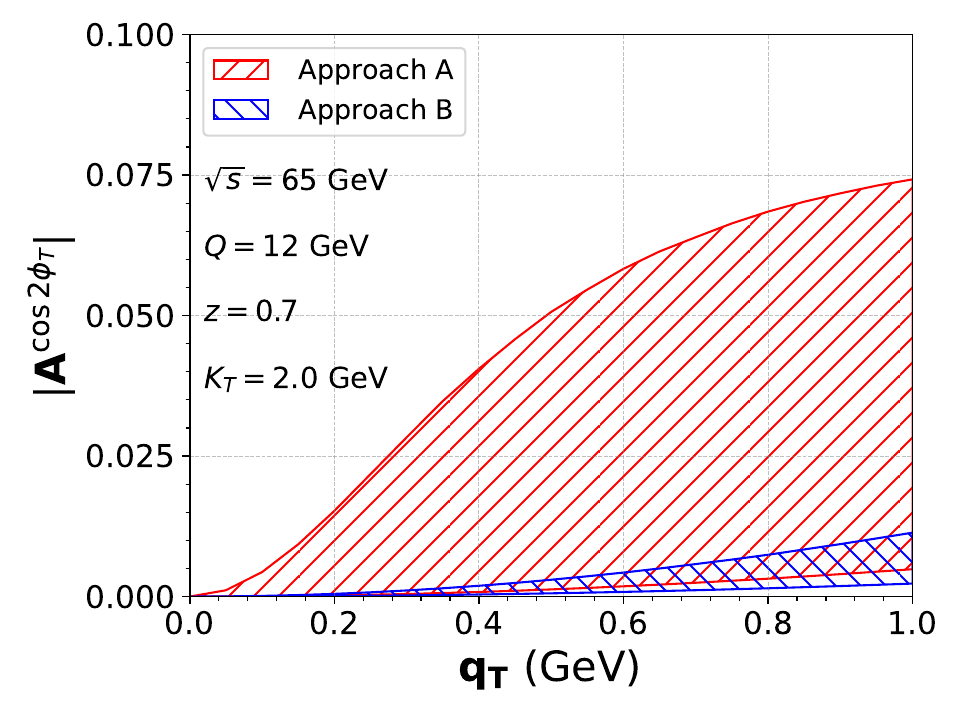}
	        \caption{}
	    \end{subfigure}
	\end{center}
\caption{\label{fig7}  $\cos2\phi_T$ azimuthal asymmetry for Approach-A and -B in $e+p\rightarrow e +J/\psi+\mathrm{jet}+X$ as function of $q_T$ at fixed values of $K_T=2.0 ~\mathrm{GeV}$, $z=0.7$, $Q=12$ GeV, in panel (a) $x=0.01$,$\sqrt{s}=140$ GeV, in panel (b) $x=0.05$,$\sqrt{s}=65$ GeV.  The bands shown in these figures are for $b_{t;lim}\{2-8\}~\mathrm{GeV}^{-1}$. The non-perturbative Sudakov factor is the same for both approaches given in Eq.~\eqref{SNPABform}.}
\end{figure}

 In this work, we consider $J/\psi$ and jet in the almost back-to-back configuration in the transverse 
plane, which allows us to use the TMD factorization.  In this kinematics,  $|q_T|\ll|K_{T}|$. We use the MSTW2008\cite{martin2009parton} set of collinear PDFs and use  NRQCD to calculate the $J/\psi$ production rate.  The contributions from both color singlet (CS)  and color octet (CO) states in the asymmetry are included.  The CMSWZ \cite{ChaoMaShaoPhysRevLett.108.242004} LDME set is adapted for the long-distance matrix elements. The kinematical region chosen in our numerical study as presented here is accessible at the upcoming EIC. We have chosen the kinematics to maximize the asymmetries. We have taken two cm  energies $\sqrt{s}=65 $ and $ 140$~ GeV.  We set $b_{T_{max}}=1.5$ GeV and have fixed the values of kinematic variables:  $z=0.7$, and $K_T=2.0$ GeV.  In Figs. \ref{fig1}-\ref{fig7}, we show the estimate for the magnitude of $\cos2\phi$ azimuthal asymmetry as a function of transverse momentum, $q_T$, which is the total transverse momentum of the outgoing $J/\psi$-jet pair and sensitive to the intrinsic transverse momentum of  the gluon in the  TMDs. We find that the asymmetry is increasing,  similar to what was observed for the process $ep\to J/\psi$ \cite{Bor:2022:EvolTMDShF}. However, the asymmetry decreases as a function of $K_T$ which we have discussed in our previous work \cite{Kishore:2022ddb}. We also found the asymmetry increases with $x$.  Here, we have chosen small values of $x$, $x=0.003$, $0.01$, where the gluons are expected to play a major role in the asymmetries. We also estimate the asymmetry in  Figs.~\ref{fig9} and \ref{fig10}, using other sets of LDMEs found in the literature.  We find that the choice of LDMEs do affect the prediction of asymmetry, as also seen in  \cite{Kishore:2022ddb}.

In Fig.~\ref{fig1}, we compare the  numerical estimates of $|\cos2\phi_T|$  asymmetry in Approach-A and Approach-B, respectively, by varying the transverse momentum $q_T$ for $x=0.01$ (a) and $x=0.003$ (b) at $\sqrt{s}=140$ GeV, $Q=4.47$ GeV, $z=0.7$, $K_T=2.0$ GeV. The range for $q_T\in[0.0-1.0]$ GeV, satisfying the condition that $|q_T|\ll|K_{T}|$. The non-perturbative Sudakov factor for Approach-A was adopted from \cite{Scarpa:2019fol}, shown in Fig.~\ref{fig:snp}(a), and for Approach-B, adopted from \cite{Bor:2022:EvolTMDShF}, shown in Fig.~\ref{fig:snp}(b).  In approach-B we show the effect of varying $b_{t;lim}=[2-8]~\mathrm{GeV}^{-1}$ as a band, whereas in approach-A we show the asymmetry for the three values of the parameter $b_{t;lim}$. For  $b_{t;lim}=2~\mathrm{GeV}^{-1}$ we don't see a significant difference depending on the approach, however, for larger $b_{t;lim}$ we start seeing a significant difference. This difference is mainly due to the large non-perturbative contribution coming in approach-A which makes the denominator smaller as compared to the numerator which results in a violation of the positivity bound for larger values of $q_T$. We have plotted only in the region where the positivity bound in satisfied. With the $S_{NP}$ adopted for approach-B, even for $b_{t;lim}=8~\mathrm{GeV}^{-1}$, the positivity bound is satisfied for a larger range of $q_T$.  In fact, as seen in  Fig. \ref{fig:snp},  $e^{-S_{NP}}$ is narrower in $b_T$ space in approach-B compared to the same in approach-A. So the effect of this factor in the larger $b_T$ region is more dominant in approach-A. In Fig. \ref{fig4}, we show the asymmetry in both approaches, however, by choosing the same non-perturbative Sudakov factor (shown in Fig. \ref{fig:snp}(b)), the asymmetry is plotted as function of $q_T$ at fixed values of $K_T=2.0 ~\mathrm{GeV}$, $\sqrt{s}=140$ GeV, $z=0.7$ and for a relatively low value of $Q$, namely, $Q=4.47$ GeV. In panel (a) we have taken $x=0.01$, and for panel (b) $x=0.003$. The asymmetry does not differ much, in this plot, which shows that for low values of $Q$, it is mainly the non-perturbative Sudakov factor that affects the asymmetry. However, this difference increases significantly at a larger scale $Q$, as seen in Fig. \ref{fig7},  plotted at $Q=12$ GeV.

\begin{figure}[H]
	\begin{center} 
	\begin{subfigure}{0.49\textwidth}
\includegraphics[height=4cm,width=6.5cm]{ 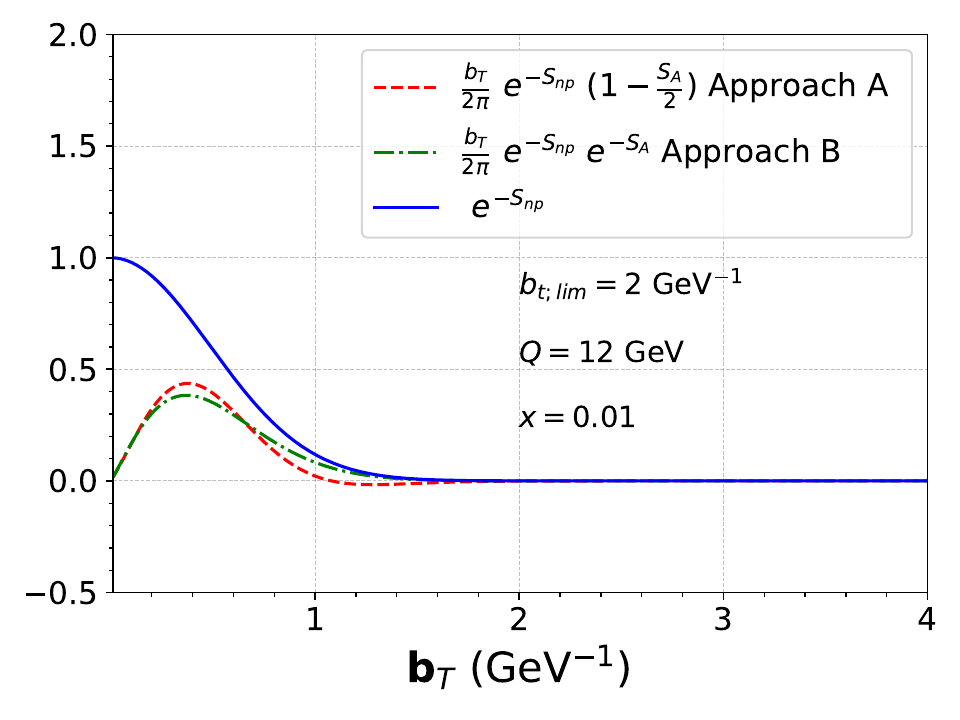}
  \caption{}
	\end{subfigure}
	    \begin{subfigure}{0.49\textwidth}
\includegraphics[height=4cm,width=6.5cm]{ 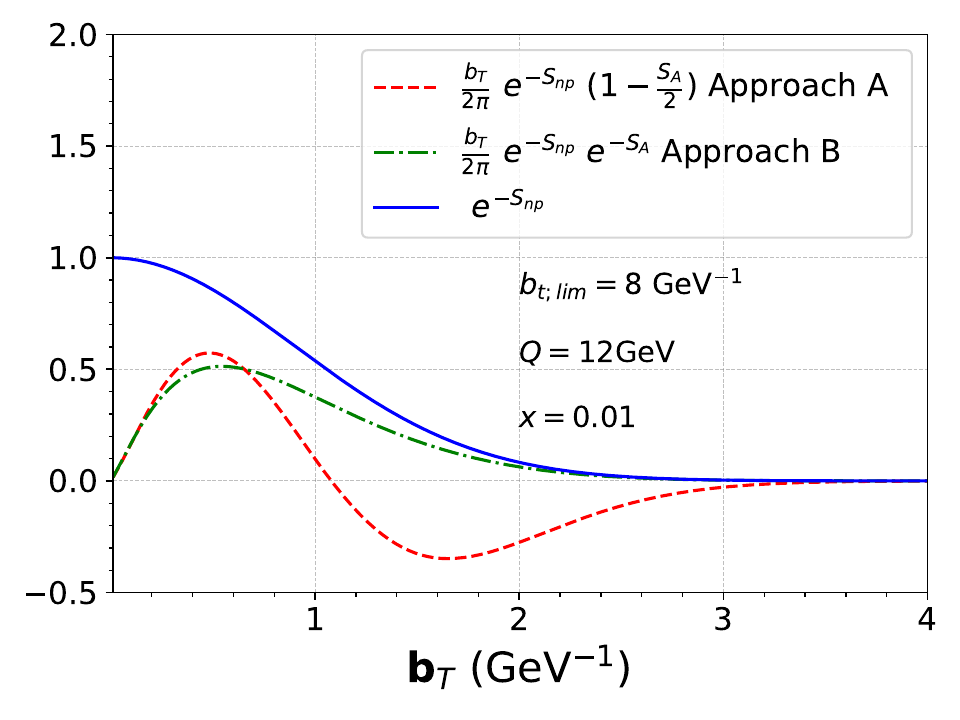}
	        \caption{}
	    \end{subfigure}
	\end{center}
\caption{\label{fig8} $b_T$ times the Sudakov factors as function of $b_T$ at $x=0.01$, and at $Q=12$ GeV for $b_{t;lim}=2$ GeV$^{-1}$ (panel (a)) and for $b_{t;lim}=8$ GeV$^{-1}$ (panel (b)). This shows the dominance of the NLO term in approach-A as the $b_T$ increases. The non-perturbative Sudakov factor is the same for both plots given in Eq.~\eqref{SNPABform}. }
\end{figure}

Fig.~\ref{fig7} illustrates the azimuthal asymmetry plotted against $q_T$ for a higher value of $Q=12$ GeV, with two different center-of-mass energies and corresponding $x$ values: $\sqrt{s}=140$ GeV, $x=0.01$ (panel (a)), and $\sqrt{s}=65$ GeV, $x=0.05$ (panel (b)). The shaded bands in the plots represent the same uncertainty on the $b_T$-width as considered in Fig.~\ref{fig7}. Observing the figures, it becomes evident that the azimuthal asymmetry is nearly identical for $b_{t;lim}=2~\mathrm{GeV}^{-1}$, in both scenarios. However, significant differences arise for the larger $b_T$-width, $b_{t;lim}=8~\mathrm{GeV}^{-1}$, between the two approaches.   The difference depicted in Fig.~\ref{fig7} shows the effect of  higher powers of the large logarithmic terms in the perturbative Sudakov kernel, in this kinematical region, which were not included in approach-A as we expanded the exponent  to a fixed order in $\alpha_s$. To explore and understand further the effect of the perturbative Sudakov factor on the asymmetry, in Fig. ~\ref{fig8} we have shown the perturbative Sudakov factor as a function of $b_T$. We have taken the same non-perturbative factor $S_{NP}$, as given in Eq.~\eqref{SNPABform}. As can be seen in the plot,  for a lower value of   $b_{t;lim}=2~\mathrm{GeV}^{-1}$, the behavior of the perturbative factor is quite similar in the two approaches, however, there is a significant difference for $b_{t;lim}=8~\mathrm{GeV}^{-1}$.  In short, the effect of the large Sudakov logarithms are more pronounced for larger scale $Q$ as well as for larger values of the $b_{t;lim}$, this is the reason for the large difference in asymmetry calculated in the two approaches.

\begin{figure}[H]
	\begin{center} 
	\begin{subfigure}{0.49\textwidth}
\includegraphics[height=4cm,width=6.5cm]{ 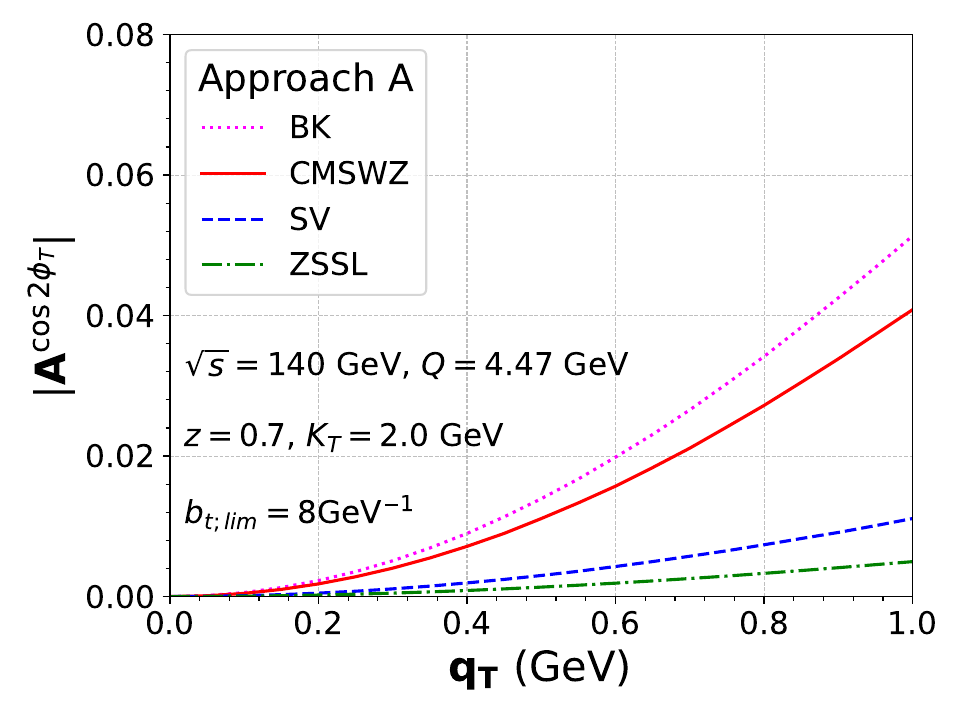}
  \caption{}
	\end{subfigure}
	    \begin{subfigure}{0.49\textwidth}
\includegraphics[height=4cm,width=6.5cm]{ 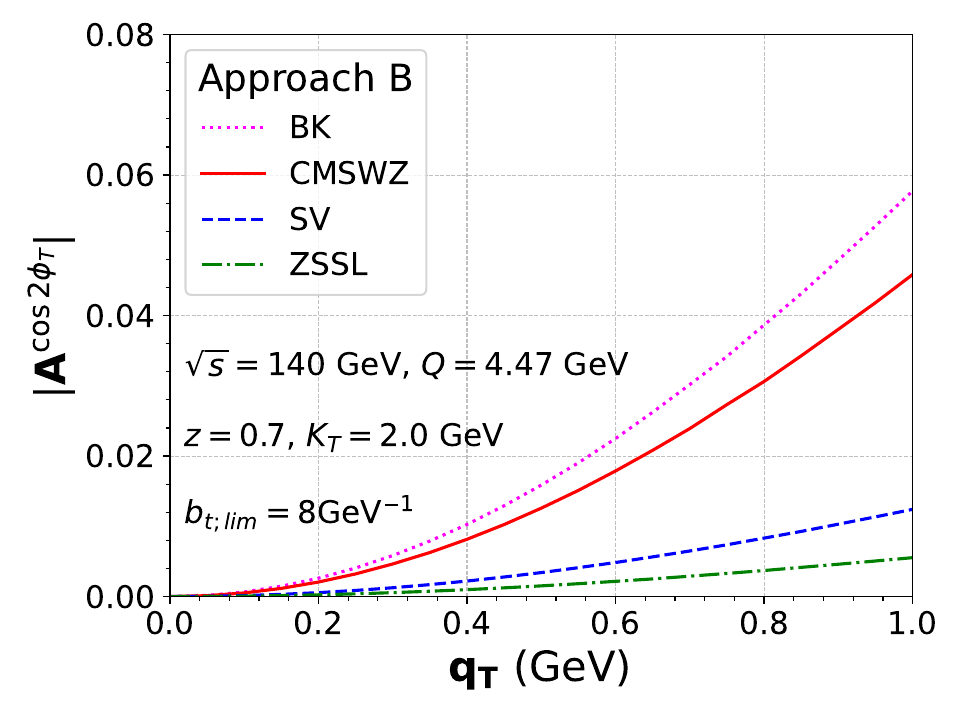}
	        \caption{}
	    \end{subfigure}
	\end{center}
\caption{\label{fig9}  $\cos2\phi_T$ azimuthal asymmetry for Approach-A (panel (a)) and approach-B (panel (b)) in $e+p\rightarrow e +J/\psi+\mathrm{jet}+X$ as a function of $q_T$ at fixed values of $K_T=2.0 ~\mathrm{GeV}$, $z=0.7$, $Q=4.47$ GeV,  $x=0.01$, and $\sqrt{s}=140$ GeV. For taking different LDME sets, namely,  CMSWZ\cite{ChaoMaShaoPhysRevLett.108.242004}, SV\cite{SharmaVitevPhysRevC.87.044905}, ZSSL\cite{ZhangSunPhysRevLett.114.092006} and BK\cite{ButenschonKniehlPhysRevLett.106.022003}, for $b_{t;lim}=8~\mathrm{GeV}^{-1}$. The non-perturbative Sudakov factor is the same for both approaches given in Eq.~(\ref{SNPABform}).}
\end{figure}   

\begin{figure}[H]
	\begin{center} 
	\begin{subfigure}{0.49\textwidth}
\includegraphics[height=4cm,width=6.5cm]{ 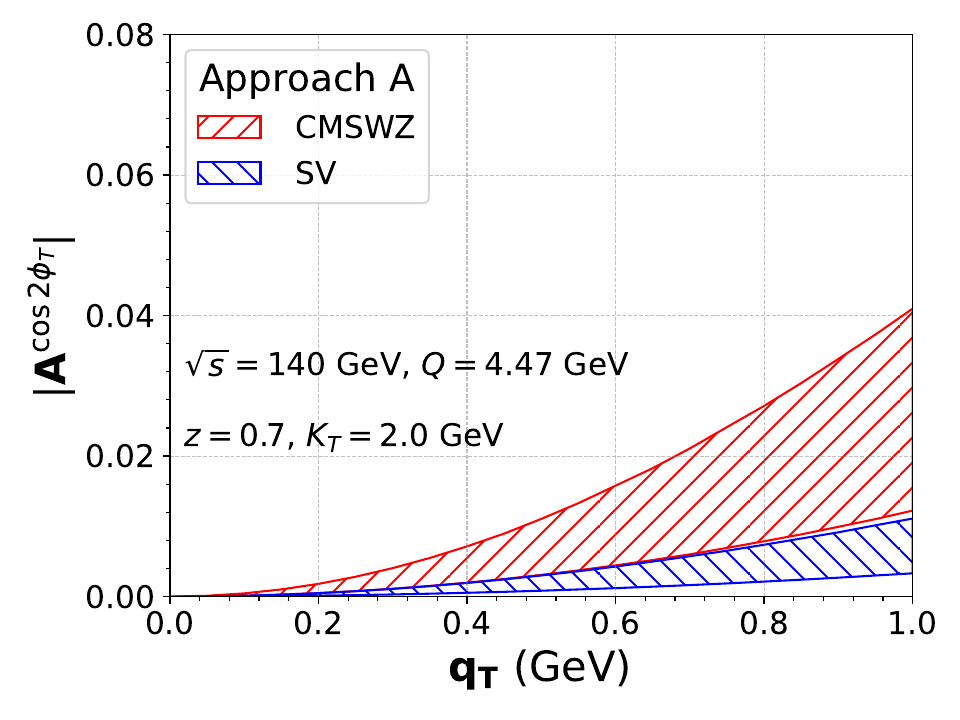}
  \caption{}
	\end{subfigure}
	    \begin{subfigure}{0.49\textwidth}
\includegraphics[height=4cm,width=6.5cm]{ 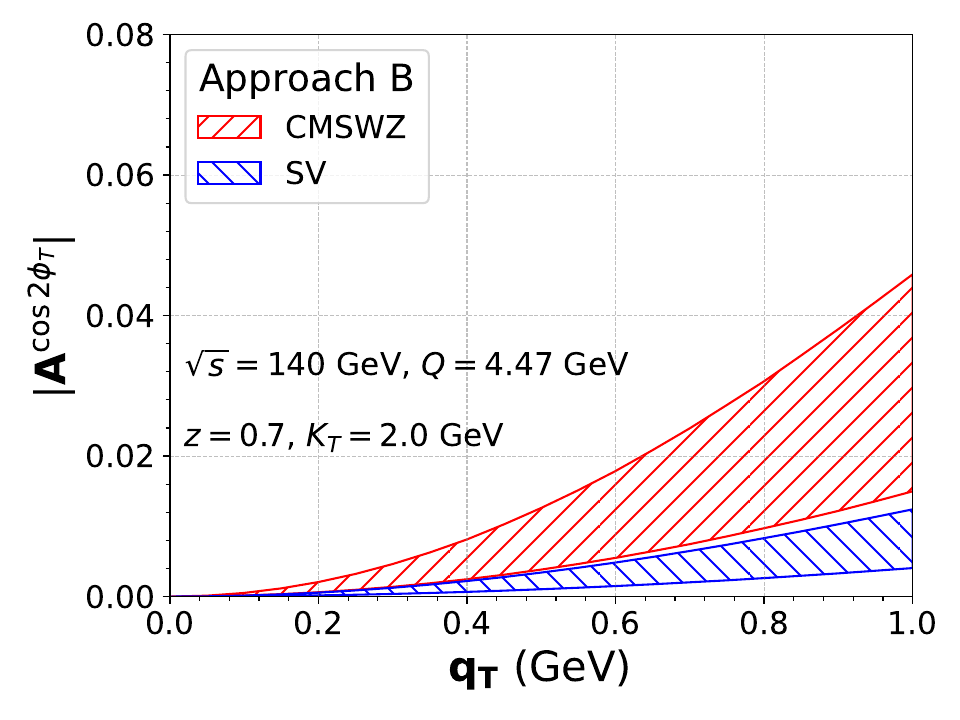}
	        \caption{}
	    \end{subfigure}
	\end{center}
\caption{\label{fig10}  $\cos2\phi_T$ azimuthal asymmetry for Approach-A (panel (a)) and approach-B (panel (b)) in $e+p\rightarrow e +J/\psi+\mathrm{jet}+X$ as a function of $q_T$ at fixed values of $K_T=2.0 ~\mathrm{GeV}$, $z=0.7$, $Q=4.47$ GeV,  $x=0.01$, and $\sqrt{s}=140$ GeV. We have taken different LDME sets, namely, CMSWZ \cite{ChaoMaShaoPhysRevLett.108.242004} and  SV \cite{SharmaVitevPhysRevC.87.044905}.   The bands shown in these figures are obtained by varying  $b_{t;lim}\{2-8\}~\mathrm{GeV}^{-1}$.  The non-perturbative Sudakov factor is the same for both approaches given in Eq.~(\ref{SNPABform}).}
\end{figure}
In the above discussion on the effect of TMD evolution on the $\cos2\phi_T$ asymmetry, we show the prediction of the asymmetry and the uncertainty (band) to the prediction, largely from the unknown non-perturbative Sudakov factor by considering $b_{t;lim}\{2-8\}$ GeV$^{-1}$, for one set of LDMEs which is CMSWZ \cite{ChaoMaShaoPhysRevLett.108.242004}.  There are several sets of LDMEs available in the literature, and as observed also in our previous work,  \cite{Kishore:2021vsm}, that the asymmetry is sensitive to the set of LDMEs used. In Fig. \ref{fig9}, we present the $\cos2\phi_T$ for approach-A and B as a function of $q_T$ for different LDME sets and for fixed $K_T=2.0 ~\mathrm{GeV}$, $z=0.7$, $Q=4.47$ GeV,  $x=0.01$, and $\sqrt{s}=140$ GeV.  
As we can see from the plots, the magnitude of the asymmetry depends on the LDME set chosen, it is larger for CMSWZ\cite{ChaoMaShaoPhysRevLett.108.242004} and BK\cite{ButenschonKniehlPhysRevLett.106.022003}, whereas for   SV\cite{SharmaVitevPhysRevC.87.044905} and ZSSL\cite{ZhangSunPhysRevLett.114.092006} it is quite suppressed. So,  the uncertainty in the LDME sets introduces an uncertainty in the asymmetry. In Fig. \ref{fig10},  we have shown the asymmetry for two sets of LDMEs, CMSWZ (red) and SV (blue) with their central values. The band shows the variation of the asymmetry when  $b_{t;lim}$ is varied in the range $\{2-8\}~\mathrm{GeV}^{-1}$. As seen in the plot, the uncertainty in the asymmetry due to the variation of the parameter also depends on the LDME set chosen.  

\section{conclusion}\label{sec5}
 In this work we discussed the effect of TMD evolution on the $\cos2\phi_T$ azimuthal asymmetry in back-to-back production of $J/\psi$ and jet in $e-p$ collision in the kinematics of the upcoming EIC. This is a promising observable to probe the linearly polarized gluon TMDs in the unpolarized proton. In our previous work, \cite{Kishore:2022ddb} we have shown that the TMD evolution reduces the magnitude of the asymmetry, or the asymmetry evolves away. Here, we investigated this further, by taking two approaches to describe the TMD evolution (in the CSS formalism) of unpolarized and linearly polarized gluon TMDs and compared their effect on the asymmetry in the kinematical range of EIC. In one  approach, we expanded the perturbative evolution kernel to a fixed order in the coupling, keeping the leading logarithmic terms,  and considered the perturbative part of TMDs up to $O(\alpha_s)$. We called this approach-A. 
In the second approach, called approach-B, we considered the leading order (LO) terms in the perturbative part of TMDs
multiplied with the evolution kernel, wherein the exponent, we kept the leading order terms.  In the literature,  the former approach is used to compare cross-sections in the TMD factorization framework to the cross-section computed in collinear factorization at some intermediate scale,  which is known as the matching procedure. The latter approach is mostly used for phenomenological studies. Moreover, we discussed the role of non-perturbative Sudakov factor $S_{NP}$, which influences the description of TMDs and thus affects the prediction of asymmetry. It is largely unconstrained for the gluon case; however, we compared two functional forms, generally chosen to be proportional to $b_T^2$. These forms differed in the width of $S_{NP}$ that regulates the contribution of perturbative tails in the higher $b_T$ region. We estimated the asymmetry functional in $q_T$ as well as the uncertainty band due to the non-perturbative factor by varying the width of $S_{NP}$ between $b_{t;lim}\{2-8\}~\mathrm{GeV}^{-1}$. We found that the asymmetry increases with $q_T$, which is sensitive to the intrinsic transverse momentum of gluon TMDs. Moreover, we observed a significant uncertainty band in the case of $S_{NP}$ with a larger width, allowing more contribution of the perturbative part in the higher $b_T$ region. We compared the effect of the perturbative part described in approach-A and -B with the same $S_{NP}$. We found both approaches have a similar influence on the asymmetry at a relatively low scale, but they differ at a large scale, where NLO terms, $\mathcal{O}(\alpha_s)$, of $f_1^g$ in approach-B dominate and suppress it. Consequently, this affects the asymmetry significantly, particularly for $b_{t;lim}=4, 8~\mathrm{GeV}^{-1}$.  We also estimated the effect of using different LDME sets on the asymmetry, and the effect of TMD evolution with different LDME sets. Overall, the asymmetry is small, not more than a few percent, in the kinematics considered,  once the TMD evolution is incorporated. Including  the shape function will improve the estimate of the asymmetry.

\section*{Acknowledgments}
This research was supported in part by the International Centre for Theoretical Sciences (ICTS) for participating in the program - International School and Workshop on Probing Hadron Structure at the
Electron-Ion Collider (code: ICTS/QEICIII2024/01). A.M. would like to thank  SERB-POWER Fellowship under file  No. SPF/2021/000102 and  Board of Research in Nuclear Sciences (BRNS), Government of India, under
sanction No. 57/14/04/2021-BRNS/57082 for financial support. M. S.  has been supported by the SERB (Science and Engineering Research Board) NPDF (National Postdoctoral Fellowship) under the File number PDF/2022/003516.  S.R. is supported by SEED grant provided by VIT University, Vellore, Grant No. SG20230031. We would like to thank M. G. Echevarria and J. Bor for the fruitful discussions. This work is partially supported by IRCC project RD/0523-IOE00I0-086.

\bibliographystyle{apsrev}
\bibliography{biblography_as.bib}
\end{document}